\documentclass[11pt]{article}
\usepackage[margin=1in]{geometry}
\usepackage[utf8]{inputenc}
\usepackage[T1]{fontenc}
\usepackage{booktabs}
\usepackage{tabularx}
\usepackage{array}
\usepackage{enumitem}
\usepackage{xcolor}
\usepackage{listings}
\usepackage[hidelinks]{hyperref}
\usepackage{xurl}
\usepackage{xltabular}
\usepackage{needspace}

\setlist[itemize]{leftmargin=1.45em,itemsep=0.18em,topsep=0.15em}
\setlist[enumerate]{leftmargin=1.45em,itemsep=0.18em,topsep=0.15em}

\newcolumntype{Y}{>{\raggedright\arraybackslash}X}

\title{The Artifact Promotion Control Model: An Implementation Case Study\\
\large Build on Target Machines vs.\ Build Once and Promote Artifacts}
\author{Vasili Gavrilov \and Enkli Ylli}
\date{First version 3 June 2026; revised 7 September 2026}

\begin{document}
\maketitle

\begin{abstract}
Artifact promotion means building the software once and moving that same built copy through the test and production environments, instead of building it again on each server. A previous treatment by the first author~\cite{ControlModel} presented it as a control model for cloud deployment, comparing build-on-target with build-once-and-promote in accessible engineering prose. This paper has two parts. Part~I restates the model in stricter form: the release object, the control domains a deployment crosses, the artifact-versus-environment identity distinction, source-control compromise as a control-domain question, and the regulatory controls whose integrity and change-control requirements the model's properties meet more directly than build-on-target does. Part~II is an anonymized implementation case study of a production web application on Amazon Web Services. A release is one command on the release engineer's machine, and the artifact's path from upload to running fleet is autonomous. Part~II reports 31 production rollouts timed stage by stage from the platform's own records, the release cadence, the service levels of the application, a control table stating which controls the implementation has and which it lacks, and the two release units the pipeline did not cover. A closing table puts each claim of Part~I to that record: what would confirm it, what the record shows, and where it stands.

\textbf{Contributions.} The parts of the model are each stated elsewhere: the build-once principle in \emph{Continuous Delivery}~\cite{ContinuousDelivery}, the build/release/run split in the Twelve-Factor App~\cite{TwelveFactor}, deployment of approved binaries only in Google's binary authorization~\cite{BAB,BSRS}, separation as a supply-chain security property~\cite{Okafor}, and the pipeline-side reading of the frameworks in NIST SP~800-204D~\cite{NIST800204D}. The paper (1)~assembles them as one control model, naming who owns each part of the path from source code to a running system and where the hand-off between them is, whatever form the built copy takes; (2)~states that injecting a secret at build time breaks artifact identity, so promotion requires secrets resolved at runtime; (3)~restates, with Zimmermann et al.'s figure, why downloading libraries at deployment time is an attack surface that grows with the number of libraries; (4)~derives from the absence of secrets in the artifact that the releasing role needs no production password; (5)~names, for seven regulatory frameworks, the deploy-time evidence each asks for and which model can produce it; (6)~reports the production record, including what the implementation does not do; and (7)~checks the model's claims against that record, claim by claim.
\end{abstract}

\section*{Preface: Why Now}

The CVE program published just over 40,000 vulnerabilities in 2024, roughly 38\% more than in 2023, the seventh consecutive record year, and submissions have grown fast enough to backlog the National Vulnerability Database~\cite{CVEStats}. Within that, software supply-chain attacks, which target the build and distribution path rather than a flaw in the running application, are a documented and growing class: the European Union Agency for Cybersecurity reports a marked rise in their number and sophistication~\cite{ENISA}, and a peer-reviewed systematization catalogues 107 attack vectors drawn from 94 real incidents~\cite{Ladisa}.

These attacks land on the build and on the path from the build to production. In March 2025 a widely consumed continuous-integration action was compromised: its version tags were repointed to code that extracted secrets from the build runner and wrote them, encoded, into workflow logs, where on public repositories anyone could read them~\cite{TJActions}. Over twenty thousand repositories depended on it. That attack hit the build step. Under a build-on-target model there is one such step per production server; under promotion there is one, whose runner has to be defended as production. The model's contribution is to confine that surface to one place with one audit trail, not to remove it.

Where a deployment failure is a safety or financial failure, in aviation, pharmaceutical manufacturing, federal and defence systems, and banking, the regulators require deployment evidence that build-on-target models produce only with substantial extra effort, if at all. Part~I makes the architectural argument and names the frameworks. Part~II documents what one implementation looks like, how long it takes, what controls it has and lacks, and what is easy to get wrong.

\section*{Terms}

The paper uses a small vocabulary, given here in plain words. A reader from compliance, audit, or programme management needs no more than this table to follow Parts~I and~II.

\begin{center}
\small
\begin{tabularx}{\textwidth}{@{}lX@{}}
\toprule
Term & Meaning in this paper \\
\midrule
Build & Turning source code into runnable software. \\
Controlled build & A build whose machine, tools, inputs, and log are recorded and access-restricted. Part~II's build is not one; see the control table. \\
Artifact & The built, deployable copy of the software: one file or image, with a version number and a fingerprint. \\
Promotion & Moving that same artifact from one environment to the next, test and then production, without building it again. \\
Environment & One complete installation of the system. Part~II has three: DEV, UAT (user-acceptance testing), and PROD (production). \\
Source control & The system that holds the source code and every change to it. A tag is a named point in that history. \\
Hash & A fingerprint computed from the exact bytes of a file; any change changes it. A cryptographic hash (SHA-256) cannot be forged for a chosen file; MD5 can. \\
Checksum & A hash used to detect accidental corruption. It proves nothing about who produced the file. \\
Signature & A hash sealed with a private key, so that it also proves who produced the file. \\
Provenance & A signed record of what went into a build: the source revision, the machine, the tools. \\
Approval & A recorded decision, by a named person, that an artifact may go to an environment. \\
Instance, fleet & One server running the application, and all of them together behind one entry point, the load balancer. \\
Content-delivery network & The service that serves the web pages and media from copies near the user. \\
Runtime configuration & The settings and passwords a running copy needs, which differ per environment. \\
Secret & A password, key, or token whose disclosure causes harm. \\
Dependency & A library the software needs. A transitive dependency is one pulled in by another library. \\
Registry & A download site for libraries or images, public or private. \\
Rolling deployment & Updating the servers a batch at a time, in place, with a health check after each batch, so the service stays up. \\
Rollback & Putting the previous artifact back. \\
Schema migration & A change to the structure of the database, applied separately from the software. \\
Snapshot & A copy of the database at a moment, from which a new database can be restored. \\
Worker & A separate background program of the system, on its own server. \\
Hook & A script the platform runs on a server at a fixed point of a deployment. \\
IAM & The cloud provider's permission system: which identity may read or write which resource. \\
Control domain & A set of assets with one owner, one audit trail, and one blast radius, the extent of harm if it is compromised. \\
\bottomrule
\end{tabularx}
\end{center}

\part*{Part I: The Control Model}

\section{The Release Object}

A common framing of the release problem is the question:
\begin{quote}
Did we build the latest code on production?
\end{quote}
A stricter framing is:
\begin{quote}
Which approved artifact is currently deployed?
\end{quote}

``Latest code'' is a property of a version-control repository and changes whenever a commit lands. An \emph{approved artifact} is a concrete object: it has a version number, a fingerprint of its bytes, a record of which source revision it was built from and when, an approval, a deployment history, and a known previous version to go back to. Engineers, project managers, testers, auditors, and stakeholders can reason about one object with these attributes; they cannot reason coherently about ``the latest code'', because each is looking at a different moving snapshot of it. In this paper the release is the artifact selected for deployment.

\section{Two Deployment Models}

Environment names vary by organization: development, integration, quality assurance, staging, user acceptance, pre-production, production, disaster recovery, customer replicas. A deployment model that depends on them is not transferable. The model here depends on one property: where the build happens.

\textbf{Model A: build on target.} Each deployment instance retrieves the source revision for the intended release, downloads every library it needs from external sites, builds on the server, and runs what it built.

\textbf{Model B: build once and promote.} One build produces an artifact identified by a version number and a fingerprint. The same artifact is stored, approved, and promoted through the environments without being built again.

\begin{lstlisting}
Model A:
    source repository -> target instance -> local build -> local deploy -> runtime

Model B:
    source repository -> build -> artifact store -> deployment -> runtime configuration -> runtime
\end{lstlisting}

Model A is fine for local development, prototypes, single-machine systems, and short-lived test hosts. Its limitation is that it puts source retrieval, library download, build, installation, and execution all on each production server, which makes the separations that controlled release needs operationally difficult. Model B makes the release boundary explicit: the production server receives a known deployable unit and does not create the release by compiling during deployment.

The models also select an environment differently. Under Model~B the environment is determined by which artifact and which runtime configuration are deployed, so one line of development remains the single source of truth. Under Model~A, teams grow a separate copy of the source per environment to carry the differences, and those copies drift: a fix applied to one is missed in another, and \emph{what is in production?} becomes a forensic exercise rather than a lookup. The same drift is documented for the configuration repositories of container platforms, where a copy per environment is named as an anti-pattern for the same reasons~\cite{Kapelonis}.

\begin{lstlisting}
Model A tendency                      Model B
  main                                  main   (single source of truth)
   |  copied by hand (drifts)            |
   +--> dev  copy                        |   one build
   +--> qa   copy   <- fixes missed      v
   +--> prod copy   <- fixes missed     artifact-X.Y.zip
                                          |  promoted unchanged
  "what is in prod?" -> forensic         +--> DEV -> UAT -> PROD
                                         "what is in prod?" -> an artifact id
\end{lstlisting}

Architects often treat the two models as interchangeable plumbing. Building the binaries once is the first principle of Humble and Farley's deployment pipeline~\cite{ContinuousDelivery}, and the control properties of the rest of Part~I depend on the build sitting off the target; they are lost, quietly, the moment it moves on.

\section{Control Domains and the Promotion Boundary}

Artifact promotion rests on a partition of responsibilities that Model~A collapses:

\begin{lstlisting}
Source repository      :  code revision history
Build subsystem        :  artifact creation
Artifact store         :  release identity
Deployment subsystem   :  controlled promotion
Runtime environment    :  configuration and secrets
Application instance   :  execution
\end{lstlisting}

Each line is a control domain: it has its own owner, its own audit trail, and its own blast radius. The claim is that source-code evolution and production change are operated in different control domains, and that the promotion boundary is the interface between them. The build may run on a release engineer's workstation or on a hosted build service; the model is defined by building the deployable unit once, giving it a stable identity, and promoting it. Whether that build is \emph{controlled} in the sense of the Terms table is a separate property, and Part~II's is not.

The model's parts are established. The binary-repository literature (JFrog Artifactory, Sonatype Nexus) names the verb \emph{promotion} and makes the artifact store the system of record for release identity. \emph{Continuous Delivery}~\cite{ContinuousDelivery} states the principle in its chapter on the deployment pipeline: build the binaries once, deploy the same way to every environment, and check the hash recorded at build time at each stage. The \emph{Twelve-Factor App}~\cite{TwelveFactor} canonicalized the build / release / run split. The control-domain reading is Google's: Binary Authorization for Borg admits to production only binaries whose signed provenance names the reviewed source and the sandboxed build, and the engineer who deploys is not the engineer who wrote the change~\cite{BAB,BSRS}. Okafor et al.\ name \emph{separation} as one of three properties a secure supply chain needs, with transparency and validity~\cite{Okafor}; the \emph{DevOps Handbook} works the segregation-of-duties case for SOX and PCI through the pipeline~\cite{DevOpsHandbook}; NIST SP~800-204D places the same trust boundaries inside the CI/CD stages~\cite{NIST800204D}; and the mechanisms the boundary depends on, fingerprinting, digital signing, and provenance, are the subject of SLSA~\cite{SLSA}, in-toto~\cite{InToto}, Sigstore~\cite{Sigstore}, and reproducible builds~\cite{ReproBuilds}. Published deployment experience reports place the practice in industry: continuous deployment at Facebook and OANDA~\cite{Savor}, the systematic review of continuous integration, delivery, and deployment by Shahin et al.~\cite{Shahin}, immutable server images~\cite{Fowler}, and GitOps as the container-side form of promotion~\cite{Beetz}; the DORA research associates continuous-delivery practice with delivery performance~\cite{Accelerate}. This paper adds nothing to those mechanisms. It draws the boundary they share, states what the two models do to it, and reports one implementation, including the admission checks that implementation lacks.

\section{Artifact Formats and the Layer of Promotion}

The form of the artifact is an implementation choice within the model: a versioned archive installed onto a prepared server, the form used in Part~II; a container image; a whole virtual-machine image; an operating-system package; a language-specific package; a disk snapshot. Each draws a different boundary between what is in the artifact and what the server is assumed to have, and the control-domain claims apply to all: each has a stable identity, each can be fingerprinted, signed, scanned for vulnerabilities, approved, and archived independently of source control, and each promotes without being built again. The choice turns on artifact size, server-preparation cost, isolation, and the frameworks below, not on the model. A team that already promotes container images, or archives through cloud storage, is using the model.

\section{Environment-Scoped Runtime Configuration}

A second separation, distinct from source/build/artifact, is between artifact identity and environment identity. The source is identical for every environment; the artifact, built bytes and deployment scripts, is identical for every environment; environment-specific runtime data is in neither. That data has two categories. \textbf{Secrets}: database passwords, API credentials, encryption keys, OAuth client secrets, mail-transport credentials, whose disclosure causes immediate harm. \textbf{Operational configuration}: database addresses, internal service addresses, storage identifiers, feature switches, region names, environment markers, which may not be secret but define the environment the artifact runs in. Both belong to the runtime environment. The implementation rule is that \emph{the artifact does not know in which environment it is running}: it starts on a prepared server and reads its configuration from one authoritative source, whether a protected file on the server, a configuration-management system, or the cloud provider's secrets service.

A further distinction is between secrets resolved at runtime on the server and secrets injected into the artifact at build time. Runtime resolution leaves each secret in one place; build-time injection copies it into the artifact, and two copies leak more easily than one. Injection also breaks artifact identity: an artifact carrying one environment's secrets is no longer the same artifact deployed elsewhere. Whether a secret lives in a protected file or a managed store is an implementation choice with a different security posture, since a managed store gives per-secret access logs, rotation, and no copy at rest on the server, and a file gives none of these; both keep the artifact promotable. Whether the secret is resolved at runtime or injected at build time is the model-level decision. Part~II resolves all secrets at runtime, into files on the server.

Model~A does not logically require secrets in source control, and a disciplined team can run it with runtime configuration correctly; the objection is structural. When every production server retrieves source, downloads libraries, builds, and installs, the boundaries among source, build settings, deployment settings, and runtime settings are easier to blur, and environment-specific files and overrides accumulate next to the source. The accidental commit is not hypothetical: a longitudinal scan of public GitHub found secrets leaking at thousands per day, and 81\% of those found were still present two weeks later~\cite{Meli}; the practices that prevent it are catalogued by Basak et al.~\cite{Basak}.

\textbf{Release authority can be separate from runtime-secret authority.} Because the artifact contains no secrets, the person who builds and promotes it needs no access to any environment's secrets. The release engineer can produce an artifact and trigger its promotion without holding any production password, while an environment administrator places the secrets in the per-environment configuration store, and a second person with environment access can roll back or redeploy without the release engineer. The separation is of duties and of where the secret is stored, not of capability: deployed code runs with the server's permissions and can read what the server can read, so whoever can deploy code can, through that code, reach the secrets. It becomes a capability boundary only when the deploy step verifies what it deploys. Under Model~A, where deploying is itself a build run with the deploying identity's access, even the separation of duties is hard to arrange.

\section{Source-Control Compromise and Control-Domain Separation}

If source control is compromised, every model requires incident response; an attacker who can alter source can affect a future build. The difference is the control path from the compromise to a production incident. Under Model~A, production servers consume the repository at deployment time, so source control is on the production execution path on every deploy and a repository compromise becomes a production compromise on the next cycle. Under Model~B, production receives an artifact built before the compromise or after it. The compromise reaches production only through a subsequent build and upload, so the exposure is one human step, with whatever review, scan, or signature check sits in that step; Part~II has none beyond the upload itself. The artifact already deployed keeps its identity, and source control is not on the deploy path. What the model adds is two control domains whose compromise reaches production without touching source control at all, the build host and the artifact store, and those two must be defended as production.

\section{Operational Properties}
\label{sec:ops}

Three properties follow from the artifact-identity boundary, and they are the ones an operations team feels first.

\textbf{Rollback is artifact selection, not rebuild.} Under Model~B, reverting a bad release redeploys a previously approved artifact still present, by identity, in the store: the bytes that ran yesterday run again. Under Model~A, rollback selects an earlier revision and rebuilds it on every server, with the same toolchain-drift exposure as the original deployment and no guarantee that today's rebuild reproduces last week's bytes. The rebuild does not reproduce even when the recipe is a container definition: of 1{,}096 container images rebuilt from their original definitions in one study, 70 reproduced the same installed library versions and 4 were identical byte for byte~\cite{DockerRepro}.

\textbf{A partial deployment is visible.} In a fleet of ten servers under Model~A, each retrieves, downloads, and builds on its own. If seven succeed and three fail, on a passing network fault, a registry outage, or a local problem, the fleet runs two versions behind one entry point and no single build log shows it. Under Model~B every server receives the same built artifact, and the platform that installs it knows which servers have it. What happens next depends on the rollout policy: under an in-place rolling update the platform halts on the failed batch and reports the split, leaving the completed batches on the new version; under an immutable or additional-batch policy the new servers are discarded and the fleet stays whole. The split is reported in the first case and prevented in the second; under Model~A it is neither.

\textbf{The production server carries less.} Model~A puts the source-control software, the build tools, and a copy of the source on every production server; each is a patching obligation, an addition to the attack surface, and a disclosure risk if the server is compromised. Model~B needs the application runtime and a deployment procedure. A related availability property follows: under Model~A a source-control outage, an expired credential, or blocked outbound network access can prevent a deployment; under Model~B the approved artifact is already present and installs without reaching any system outside the cloud account.

\section{Regulatory Controls the Properties Serve}

No framework below requires build-once or promotion. Each requires an outcome: integrity verification, change control with segregation of duties, traceability to a validated executable, or no unapproved software inside a boundary. Promotion is one means, and the entries state, for each framework, the deploy-time evidence it asks for and why build-on-target cannot supply it. The frameworks are written in the vocabulary of auditors, but the obligations land on the engineers who operate the deployment. The research literature on continuous delivery in safety-critical settings treats the same requirement from the process side, as the obligation to re-verify every delivered build against the safety analysis~\cite{VostWagner}.

\textit{FedRAMP and NIST SP 800-53.} FedRAMP authorizes cloud services that handle U.S.\ federal data on the controls of SP~800-53~\cite{NIST80053}. SI-7 (``Software, Firmware, and Information Integrity'') requires integrity-verification tools and a response to a failed check; the Moderate baseline adds SI-7(1), checks at start-up or on a schedule, and the High baseline adds SI-7(15), authentication of code before installation. CM-5 restricts who may change what, and CM-14 requires signed components. Promotion supplies the reference value these checks need: a fingerprint recorded once at build time, stored with the release, and compared before a server accepts the artifact. Under Model~A each server builds its own bytes and leaves nothing fixed to check against.

\textit{Sarbanes--Oxley Section 404.} Section~404~\cite{SOX} requires management of a public company to assess, and its auditor to attest, internal control over financial reporting. Segregation of duties between code author and production deployer is not in the statute; it arrives through the auditor's tests of change management, which ask who can move a change into production and whether an approval exists for each. Where a production server may pull and compile source without a gate, every author holds production-execution rights. Promotion makes the separation mechanical, the resolution the \emph{DevOps Handbook} reaches for the same controls~\cite{DevOpsHandbook}: the commit, the approved artifact, and the promotion step each carry their own owner and audit trail.

\textit{FFIEC IT Examination Handbook.} The handbook the U.S.\ banking regulators examine against~\cite{FFIEC} expects a change record with a request, approvals, test evidence, a back-out plan, and a post-implementation review. Promotion supplies the artifact identity, the deploy timestamp, and the rollback record in that change record; the request, test evidence, and approval come from the change-management system in either model. Model~A supplies the identity only as a history reconstructed from per-server build logs.

\textit{RTCA DO-178C.} The civil-aviation software guidance~\cite{DO178C}, recognized through FAA AC~20-115D, identifies one executable object code by part number in the Software Configuration Index, records in the environment configuration index the compiler, linker, and options that produced it, and binds the verification record to that executable. A deployment that recompiles on each target ships bytes the index does not name. A workstation build with no recorded toolchain, as in Part~II, would not meet it either.

\textit{FDA 21 CFR Part 11.} The rule on electronic records in pharmaceutical, biotech, and medical-device processes~\cite{Part11} requires validated systems (11.10(a)) and secure, time-stamped audit trails (11.10(e)); FDA's software-validation guidance ties validation to the installed version. The validated unit is therefore the artifact with the validation runs executed against those exact bytes. A per-server compile produces bytes no validation record describes.

\textit{HIPAA Security Rule.} The Security Rule's integrity standard, 45~CFR~164.312(c)~\cite{HIPAA}, protects the health data itself. Software change control enters through the risk-management requirement of 164.308(a)(1): a covered entity that lists uncontrolled deployment as a risk needs a control, and a fingerprinted, promoted artifact is one.

\textit{Department of Defense Impact Levels IL-4 to IL-6.} The DoD Cloud Computing Security Requirements Guide~\cite{CCSRG} defines the levels. IL-4 and IL-5 restrict internet access to the DoD boundary; IL-6 has none. A model that fetches source and libraries at deploy time depends on an approved mirror inside each boundary; a self-contained artifact crosses it with one transfer. The Department's reference design implements this as a curated registry of approved, hardened container images, re-scanned inside each enclave~\cite{DoDDevSecOps}.

Promotion does not by itself produce compliance with any of these; compliance is a larger system property. The point of the enumeration is that the boundaries the frameworks require are the boundaries the model establishes, and that a team expecting to be evaluated under any of them is better off drawing them early than under audit pressure.

\section{Common Objections}

\textbf{``Old artifacts ship vulnerabilities.''} Old artifacts should not be promoted indefinitely, and a bundled library receives no upstream fix until the artifact is rebuilt. Vulnerability response is cleaner under promotion: rebuild once, identify, scan, approve, promote. The alternative is coordinating source pulls and local builds across a fleet with no way to verify that all servers built equivalent bytes.

\textbf{``This is overengineering.''} An artifact store, a versioned package, a deployment procedure, and an environment-scoped configuration set establish the boundary. The complexity of larger implementations belongs to the systems they support.

\textbf{``Source on the server helps debugging.''} Production debugging uses logs, metrics, and controlled diagnostic packages. A copy of the source on the production server is not a release process.

\textbf{``Promotion means committing to one CI vendor.''} The build can run on a workstation, Jenkins, GitHub Actions, GitLab CI, Azure Pipelines, AWS CodeBuild, or anything else. The model is defined by building once, giving the artifact an identity, and promoting it. Part~II builds on a workstation and deploys through an AWS service used as a script runner; nothing in the model requires either.

\part*{Part II: Implementation Case Study}

\section{Setting}

The system is a production web application: a server side answering requests from the browser, web pages served from a content-delivery network, and a relational database run by the cloud provider. The programming language is immaterial to what follows. The fleet is interchangeable application servers behind one load balancer; the source is one shared line of development with a tag per release. The infrastructure is on Amazon Web Services, and the specific services (S3, EventBridge, CodeBuild, Elastic Beanstalk, CloudFront) are named because a recipe is useful only if concrete; equivalents exist elsewhere. Withheld is anything that would grant access or identify the deployment: account identifiers, permission roles, storage names, hostnames. All dates and times are UTC.

There are three environments. The pipeline was built on DEV in June 2026 and stood up on PROD on 27 July and on UAT on 29 July 2026; DEV was retired on 11 August. Before the pipeline, each environment was installed by a hand-run script on a single server. Through the period reported here UAT carried the live users while PROD was being brought up, so UAT is the environment with users behind its figures and PROD the one whose rollouts are timed. Both fleets are two servers in two availability zones, updated in place one server at a time with a health check after each. The system was not subject to any framework named in Part~I, no audit was performed, and no compliance is claimed.

\section{Materials and Methods}

\subsection{The deployment pipeline}

A release is one command on the release engineer's machine: a build that produces a versioned zip file, then an upload of that file to the environment's release store. The upload is the only action on the artifact's path; what the command does besides is in the section on uncovered units. The listing below is the path the upload sets off. A non-technical reader can take it as: the upload is noticed; a script publishes the web pages and hands the artifact to the platform; the platform installs it on each server in turn; each server fetches its own settings as it starts.

\begin{lstlisting}
Release engineer (local):
    release build  produces  artifact-<VER>.zip   (built locally; no cloud build)
    aws s3 cp      artifact-<VER>.zip -> s3://<release-store-for-env>
        |
        | S3 ObjectCreated event
        v
Amazon EventBridge rule (one per environment)
        | fires StartBuild on the CodeBuild project
        v
AWS CodeBuild project (source = NO_SOURCE)
        | downloads the artifact from S3
        | runs the deployment script (buildspec):
        |     - sync static assets to the CloudFront origin bucket (with delete)
        |     - create a CloudFront invalidation
        |     - assemble the Elastic Beanstalk source bundle from the artifact
        |     - upload bundle, CreateApplicationVersion, UpdateEnvironment
        |     - poll until the environment is Ready; verify the status endpoint
        v
AWS Elastic Beanstalk (in-place rolling update, one server per batch)
        | installs the new application version on each server in turn
        v
Per server, at predeploy (.platform hook):
        | the hook probes each candidate config store in S3;
        |   the server's IAM role may read exactly one, its environment's
        | pulls the config files to local paths, sets owner and mode
        v
Application restarts and reads its config on startup
\end{lstlisting}

\subsection{Trust boundary}

Whoever can write to an environment's release store can deploy any code to that environment: the script deploys whatever lands, there is no approval step between upload and production, and the deployed code, including the predeploy hook, runs with the server's permissions. That write permission is therefore the production change permission and has to be granted as one. The build records an MD5 checksum beside the zip, and the project's release archive records a SHA-256; no stage of the pipeline verifies either before installation, and both are written by the identity that writes the zip, so they detect corruption in transit and identify the artifact, and they are not tamper evidence. The artifact is not signed, not scanned, and generates no provenance; the build is on a workstation whose machine and tools no system records. The release and configuration stores are versioned and encrypted at rest, so an upload under an existing name keeps the previous bytes retrievable; the uploader's identity is not recorded by the pipeline; S3 data-event logging, which would record it, is a separate account setting. The decision to release is taken by the team and recorded in the project's development forum, and each release is tagged in source control and archived to the project's release page with a SHA-256, under the releasing account; both records are outside the pipeline, which checks for neither, and nothing checks that the artifact uploaded to PROD is the one that ran on UAT. Each missing check is a one-step addition. The control table at the end of Methods states them together.

\subsection{The service named CodeBuild runs the deployment and builds nothing}

The CodeBuild project's source is \texttt{NO\_SOURCE}, and the product name works against the reader: no compilation occurs in this stage. The artifact was built before the upload. CodeBuild is used as a script runner, started by the upload and permitted only what the deployment needs: it downloads the artifact and runs the deployment script; it retrieves no source and downloads no libraries. Engineers reading the architecture for the first time assume the build happens here and look in the wrong place when diagnosing a failure.

The script publishes the web pages before it hands the artifact to the platform, deliberately: the new pages reference asset names the old pages do not, and the reverse order would leave a window in which new server code serves pages whose assets are missing. The cost is the mirror window: on every rollout the new pages precede the new server code by the six minutes of the rollout, on a rollback the old pages precede the old code, and a script that fails after the publish step, which the three refused runs in Results did, leaves the newer pages against the older code until the next successful run. Content-hashed asset names with the entry pages published last would close both windows; the implementation has not done this.

\subsection{Libraries}

The application's third-party libraries are committed to the source repository and reviewed as code; nothing is downloaded at build time or at deploy time from any registry. Downloading libraries at deploy time fails in two ways that need no adversary: a maintainer withdraws a published version and a download that worked last week no longer does, and two builds of the same source weeks apart pick different library versions and produce different bytes. The third is adversarial. Every public registry is a code-injection surface, and the risk grows with the number of transitive dependencies: installing an average npm package implicitly trusts about 80 others~\cite{Zimmermann}, and malicious packages are a recurring, catalogued class of attack~\cite{Ohm,Ladisa}. Under Model~A that download happens at deploy time, on or near production, on every rollout. Under Model~B it happens at most once, at build time, away from production, and here not at all. The cost is that a bundled library receives no upstream fix until someone updates it and rebuilds; the implementation keeps no software bill of materials and scans nothing, so that obligation rests on the reviewers.

\subsection{Environment identity via IAM scoping}

Each server must learn which environment it is in so as to fetch the right settings. The platform does carry an environment name, and a setting could be placed on the server by hand; the point of the design is stronger than either: a server cannot read another environment's secrets whatever it believes about itself, because each environment's servers are permitted to read exactly one configuration store. At start-up the hook tries each candidate store in a fixed order and takes the first the permission system allows; if none is allowed it refuses to start. It does not detect two, which would mean an over-granted policy; the fetched settings carry an environment marker that the hook compares with the store's name, and that comparison catches a file uploaded to the wrong store, not an over-granted policy. Each start-up also produces denied probes on the other environments' stores, which a security monitor will flag unless allow-listed.

\subsection{Per-environment configuration}

The per-environment configuration is a few files: the application's settings, its logging settings, and the web server's settings. They are uploaded once per environment from a secured machine and never written by the pipeline. The artifact is identical byte for byte across environments; the environment-specific data, including database and mail credentials, lives in the configuration stores, encrypted and access-controlled, and, once fetched, in files on each server's disk, readable by the service account. Because the artifact carries no secrets and the stores are written separately, triggering a deployment needs only the right to upload an artifact, so the role that releases and the role that holds production secrets can be different people. In this implementation the roles are three: the developers, the release engineer (the first author), who builds and uploads, and the systems operator (the second author), who administers the environments and the secret stores. The release engineer also holds environment access, which the release command uses for the database and worker steps, so the separation Part~I describes is one of duties, recorded in the forum decision, and not one the pipeline enforces.

\subsection{Control table}

The table states, for the controls an auditor would ask about, what this implementation produces, which identity produces it, and where the answer is ``not implemented''.

\begin{center}
\begin{tabularx}{\textwidth}{@{}lYY@{}}
\toprule
Control & Evidence the pipeline produces & Status \\
\midrule
Artifact identity & Version number in the file name; MD5 beside the zip; SHA-256 in the release archive; platform version label & Implemented, not verified at deploy \\
Integrity check at deploy & None & Not implemented \\
Signature, provenance & None; workstation build & Not implemented \\
Vulnerability scan, bill of materials & None & Not implemented \\
Approval & A team decision recorded in the development forum before each release & Recorded outside the pipeline; not checked by it \\
Segregation of duties & Developers, a release engineer, and a systems operator; the release engineer also holds environment access & By role and record; not enforced by the pipeline \\
Same artifact in UAT and PROD & By procedure: the same zip, taken from the release archive by tag & Not checked at deploy \\
Deployment record & Runner build history; platform event log per server & Implemented; retention not set beyond platform defaults \\
Releaser identity & The tag and the release-archive entry carry the releasing account & Recorded in source control; the cloud upload itself is not attributed \\
Artifact retention & Versioned release store; project release archive & Implemented; no stated lifecycle \\
Rollback & Re-upload of a previous artifact from the archive; the platform's own version-label rollback exists and is unused & Implemented, not exercised \\
Test evidence tied to an artifact & UAT deployment is the only pre-production exposure; its outcome is not logged against the artifact & Not implemented \\
Build environment recorded & No & Not implemented \\
\bottomrule
\end{tabularx}
\end{center}

\section{Results}

The application is about 121{,}000 lines of server-side source and 88{,}000 of browser-side source; the artifact is 10~MB.

\subsection{The first autonomous deployment}

The first autonomous end-to-end deployment, on the one-server DEV environment, took about 90 seconds from upload to the running application reporting its new version: about 5 s for the upload to be noticed, 30 s for the script's own work, 28 s to install on the one server, and 5 s for the health check.

\subsection{Rolling deployment on the production fleet}

The runner and the platform each keep their own timestamps, so every PROD deployment can be timed after the fact: the runner's build history gives its start and end, and the platform's event log gives the start of the environment update, each server's completion, and the moment the new version is reported deployed. The table pairs them for the 31 rollouts between 28 July and 3 September 2026 in which both records are complete; the individual rows are in the Appendix. The policy is the platform's in-place rolling update with a batch of one server and its default health-check thresholds, which were not tuned.

\begin{center}
\begin{tabularx}{\textwidth}{@{}Xrr@{}}
\toprule
Stage & median & range \\
\midrule
Script start to the platform beginning the update (web pages published, package assembled, version registered) & 27~s & 11 to 39~s \\
First server: bundle downloaded, settings fetched, application started and answering & 36~s & 36 to 38~s \\
Second server, from the first server's completion (health check, then the same installation) & 170~s & 150 to 172~s \\
From the second server's completion to the platform reporting the new version deployed & 138~s & 127 to 149~s \\
Platform update, start to reported deployed & 344~s & 319 to 360~s \\
Script, start to end (the difference from the row above is the script's own polling and status check) & 398~s & 352 to 411~s \\
\bottomrule
\end{tabularx}
\end{center}

The 26 rollouts on UAT differ only in the first row, 72~s against 27~s, from more web pages to publish; the platform stages match to within a few seconds. The first server's row, 36 to 38~s across 31 rollouts, is the whole of what the artifact costs a server: download, settings, start. The second server's row and the settle row are the same health gate twice, the load balancer's healthy threshold times its check interval plus the platform's own reporting cadence, and they contain a second 36-second installation. A faster rollout comes from those thresholds and the batch size, and a slower application start would lengthen the gate. The spread is 15\% of the median end to end, so a deploy that runs long is a signal in itself. Under Model~A the 36-second row would also carry a source retrieval, a library download, and a build per server, with their variation, and would have to fit the platform's per-server command timeout; the gates around it would be unchanged.

The stand-up day is not in the table. On 27 July the first PROD rollout failed on one of the two servers; the platform halted the update, reported the split, and left that server unhealthy while the other served, so the fleet was one server for the five minutes until the second attempt, which the platform's log shows completing on both servers while the runner's own verification reported the run as failed. Three later runs, on 14, 19, and 24 August, ended within 20 seconds of starting because a second artifact was uploaded while a rollout was still in progress and the platform refused a new update until the environment was ready; each had already published the newer web pages, and the remedy was to upload again after completion. In all, 36 runs: 31 complete, 2 on the stand-up day, 3 refused.

\subsection{Release cadence}

The project carries 63 tagged releases since its first in March 2026. In the 29 days from 27 July to 24 August 2026, the first four weeks after PROD ran the pipeline, 22 were built once on the release engineer's machine and uploaded. A promotion to the next environment is the upload of the same zip file to that environment's store; nothing is built again and no second tag is cut. The platform's version label joins the artifact version to the script run's number, so the history can tell two uploads of the same file apart while the file is the same.

\subsection{Service levels of the application}

The figures below are the application's, on UAT, over the 92 days from 1 May to 1 August 2026, taken from access logs and the load balancer's metrics. For 89 of those days UAT was installed by the previous, hand-run path; the pipeline reached it on 29 July. They therefore describe the deployed application and say nothing about the model; they are reported because the same application ran on under the model with no change in service, and because the connection budget below rests on them. Health checks and internet scanners are 96\% of the log, so the other rows are computed over user requests only, identified by their path.

\begin{center}
\begin{tabularx}{\textwidth}{@{}lX@{}}
\toprule
Requests in 92 days & 504{,}616 health checks, about 40{,}000 scanner probes, 20{,}851 user requests \\
Response time, user requests & 95 of 100 answered within 64~ms; 99 of 100 within 316 to 392~ms, the range across measurement days \\
Errors & 28 server errors (HTTP 5xx) in 20{,}851 requests, 0.13\% \\
Largest load carried & 124 users with a session open in the same five minutes, 15.6\% peak database processor use, 7 database connections open of a limit of 60 \\
\bottomrule
\end{tabularx}
\end{center}

The database is the smallest size the provider offers, so these are a floor, not a capacity. The resource that would run out first is database connections, and its budget is set by the number of servers rather than by the number of users: each server holds a pool that may grow to a ceiling of 22 connections, and the budget must assume every pool at its ceiling:

\begin{lstlisting}
servers x pool ceiling  +  1 worker  +  3 internal  <=  0.8 x database limit
\end{lstlisting}

On UAT it is 2 x 22 + 1 + 3 = 48, which is 0.8 x 60, so the budget is exactly spent. This is a computation; no rollout in the record exhausted connections, and the peak observed was 7. Under the in-place rolling policy used here old and new servers do not coexist, so a deploy does not double the demand; what does is a second fleet, an added server, a rollback that leaves a previous tier running, or a policy that adds a batch, and the one exhaustion the team saw came from a previous tier kept alive for rollback during the migration to the pipeline. When the budget is short the fixes, in order of cost, are a smaller pool ceiling, closing the pool on shutdown, a connection proxy, and a larger database; an added server consumes the scarce resource rather than relieving it.

\subsection{What the promotion boundary did not cover}

A release of this system is three units: the artifact, the database changes it may carry, and a background worker program that runs on a separate permanent server outside the fleet and obtains the new release by downloading the same zip from the release store. The pipeline promoted the first. The other two were runbook steps, and an application deploy left the worker running the previous release's code until someone restarted it. The failure is silent: the old worker admits or refuses work by the previous release's rules, and the symptom is that nothing happens, not an error. On 19 August 2026 the release command was extended to drive all three units: it uploads the artifact, applies outstanding database changes only if the environment's recorded database version is behind, after a database snapshot on PROD, and restarts the worker on the new release every time, unconditionally, because the alternative is a hand-kept list of what the worker loads, and such a list rots. The restart is graceful: the job in hand finishes first. The order has a hazard of its own: the upload returns in seconds and the rollout runs for six minutes, so the database change is applied while one server runs each release. The rule that makes this safe is that every database change must be tolerated by the release before it, and it is a rule, not a check.

The second hazard is of the same kind. The settings are fetched by the hook when the platform deploys to a server; editing the store changes nothing already running, and restarting the application does not fetch again. A settings change reaches the fleet by a redeploy of the current version, which re-runs the hook, or by replacing the servers, not by a restart.

\section{The Model's Claims Against the Record}

Part~I makes claims about what the model gives a deployment. The table states each as something the record could confirm or refute, and what the 36 runs, the control table, and the pipeline's construction show. The first block establishes what the system is, from its logs and its repository; the second is the promotion boundary itself, the third the operational properties of Section~\ref{sec:ops}, the fourth the controls, including the three properties Okafor et al.\ name for a secure supply chain~\cite{Okafor}.

\needspace{12\baselineskip}
{\small
\begin{xltabular}{\textwidth}{@{}p{0.2\textwidth}YYp{0.15\textwidth}@{}}
\toprule
Claim & What would confirm it & What the record shows & Standing \\
\midrule
\endfirsthead
\toprule
Claim & What would confirm it & What the record shows & Standing \\
\midrule
\endhead
\bottomrule
\endlastfoot
\multicolumn{4}{@{}l}{\emph{The system}} \\
\midrule
A production system with concurrent users & User traffic and its peaks in the platform's own records & 20{,}851 user requests in 92 days; load arrives in peaks, and on 6 August 2026, for example, 124 users held sessions in the same five minutes, at 15.6\% peak database processor use; 95 of 100 requests answered within 64~ms; 28 server errors in 20{,}851 requests & Measured \\
Released at production cadence & Tagged releases and their dates & 63 tagged releases since March 2026, 28 of them between 27 July and 3 September under the pipeline; five developers and two subject-matter experts; 1{,}736 commits in ten months; 163{,}000 source lines excluding vendored code, 47{,}600 of them end-to-end tests & Measured from the repository \\
The mechanism is small & The pipeline's definition counted in lines & The artifact's path is 571 lines in four files: the infrastructure template (261), the deployment script (147), and two server hooks (163); the release command is 613 lines in two files; standing up an environment is one 682-line script & Measured \\
\midrule
\multicolumn{4}{@{}l}{\emph{The boundary}} \\
\midrule
Artifact identity: the same bytes reach every environment & A digest recorded at build and checked on each server & One zip per release, uploaded unchanged to each store; MD5 beside the zip and SHA-256 in the release archive; no stage checks either & Holds by procedure; unchecked by the pipeline \\
The version that was uploaded is the version that runs & The rollout verified against the artifact's identity before the run is called good & The runner waits for the platform and fails the run if the environment does not report the new version; after each rollout the application's status endpoint, the CDN's version file and the platform's version label are compared & Holds for the version label; the bytes are not compared \\
A shipped identity cannot be reused, and a command reaches one environment & A release refused when its version already exists; no command spanning two environments & The release command refuses a version whose tag is already on the origin, checked before the upload, which is the point of no return; it takes one environment per run and requires the name typed back for UAT and PROD & Holds by the release command \\
The environments are the same infrastructure & Both stood up from one definition, with a check for drift & UAT and PROD are stood up from one template; a parity script diffs their settings, roles, platform, version, instance count and secret residue against PROD as reference & Holds by template; the record does not say when the check last ran \\
Promotion is an upload, not a rebuild & Releases promoted with no second build and no second tag & 22 releases in 29 days built once; each promotion the upload of the same zip; the platform's label joins artifact version to run number & Holds \\
No secrets in the artifact; settings from a per-environment store & The artifact identical across environments; each server able to read one store & Byte-identical artifact in UAT and PROD; each server's role reads exactly one store, and a server that can read none refuses to start & Holds by construction \\
A server learns its environment from what it may read & A wrong-store upload caught; an over-granted policy caught & The settings' environment marker is compared with the store's name, which catches the first; nothing detects two readable stores & Partial \\
Source control holds no environment data & No settings or secrets in the repository or the artifact & Settings uploaded once per environment from a secured machine, never written by the pipeline; libraries committed, settings not & Holds by procedure \\
No source checkout, dependency download, or build on the production host & The server's work during a rollout limited to download, settings, start & 36 to 38~s per server across 31 rollouts for download, settings and start; the runner's source is \texttt{NO\_SOURCE}; libraries are committed, none downloaded & Holds; no host inventory was taken \\
The boundary encloses everything a release changes & Every release unit carried by the pipeline & Three units; the pipeline carried one, and an application deploy left the worker on the previous release until 19 August, when the release command was extended to all three & Failed, then corrected \\
A settings change reaches the running fleet & A store edit followed by the fleet running the new settings & The hook fetches at deploy only; an edit changes nothing running, and a restart does not fetch again; a redeploy or replaced servers are needed & Holds only through a redeploy \\
\midrule
\multicolumn{4}{@{}l}{\emph{Operation}} \\
\midrule
The releasing role needs no production password & A release performed with the upload right alone & The upload right deploys the artifact; the release command also applies database changes and restarts the worker, which use environment access & Holds for the artifact; not for the whole release \\
Rollback is artifact selection & A rollback performed by re-uploading a previous artifact & The operation exists and is the same as a deploy; none was needed in the window & Not exercised \\
A database change is reversible & A snapshot before every migration and a restore exercised & On UAT and PROD the release command takes a manual snapshot, waits for it, and refuses to migrate without one; a restore creates a new instance rather than overwriting; no restore has been exercised & Partial: snapshot yes, restore untested \\
A deploy fails closed & A missing permission producing a failed run, not a partial one & A runner-role permission gap fails the run, with the platform's message naming the missing action; observed while the environments were stood up & Observed \\
A rollout does not double the database's load & Old and new servers not coexisting & In-place rolling, one server at a time; the one connection exhaustion seen came from a previous tier kept alive for rollback during the migration to the pipeline, not from a rollout & Observed once, outside a rollout \\
A partial deployment is visible & A failed batch reported by the platform with the fleet's state & 27 July: the update halted on one server, the platform reported the split, the other server served & Observed once \\
Rollout duration is itself a signal & A narrow spread across rollouts & 31 rollouts end to end in 352 to 411~s, a spread of 15\% of the median; the artifact's own term 36 to 38~s & Holds \\
Concurrent releases are refused, not merged & A second upload during a rollout rejected & Three runs, 14, 19 and 24 August, ended within 20~s because the platform refused an update while one was in progress; each had already published the newer web pages & Refused as claimed; the page-first window is a defect \\
Installation needs no system outside the cloud account & A rollout with no external fetch & The path is store, event rule, runner, platform; nothing is fetched from a registry or a source host & Holds by design; no outage was tested \\
Models diverge at fleet scale and on rollback (Part~I) & The same system measured under both models & Model~B measured; no Model~A measurement of this system exists & Not tested \\
\midrule
\multicolumn{4}{@{}l}{\emph{Controls}} \\
\midrule
Separation (Okafor et al.) & Distinct identities for build, release and secrets, enforced by permissions & Three roles by record; the release engineer also holds environment access & Partial; by record, not enforced \\
Transparency (Okafor et al.) & Every step attributable to an identity and logged, and kept & Runner history and platform event log per server, and each server's hook log; application logs kept 180 days, hook and engine logs 14; the upload is not attributed & Partial \\
Validity (Okafor et al.) & Signature or provenance verified before installation & None: workstation build, no signature, no scan, no check at deploy & Absent at this stage; to be added \\
Attack surface confined to one build & No dependency download at deploy; the build recorded & No download at deploy; the build is on a workstation that no system records & Deploy-time surface closed; build-time surface uncontrolled \\
Deploy-time evidence for the frameworks of Part~I & A change record, an approval, and an integrity check, each tied to the artifact & A deployment record per server and an approval in the forum, neither tied to a digest; no integrity check; no audit was performed & Partial; no compliance claimed \\
\end{xltabular}}

Of thirty rows, the first three establish the system and the mechanism by measurement. Of the twenty-seven claims, fourteen hold on the record, four of them by construction of the permissions or by procedure rather than by any check the pipeline runs, and one for the version label rather than the bytes. Seven are partial, and the three a supply-chain reviewer asks about first, validity, an attributed upload, and a recorded build, are absent. One failed in the window and was corrected, one was never exercised, and the comparison with build-on-target that motivates Part~I is not tested by this record and cannot be with this system: build-on-target was designed for it in the working notes and never installed.

\section{Discussion}

Part~I predicts that the models diverge at fleet scale and on rollback. The record fixes the Model~B side of that comparison and measures nothing on the Model~A side. Across 31 rollouts the artifact's cost on a server is 36 to 38 seconds, and the rest of the six minutes is the platform's health gating, which either model would pay; what Model~A would add is a build inside that 36-second term on every server, with its variation and its dependence on external sites.

A rollback here is the same operation as a deploy: the promotion command uploads the previously promoted artifact, from the project's release archive, and the script and the platform run the same stages, web pages included, so the table is also its expected duration. None was needed in the window. The platform's own rollback, re-selecting a retained version label, is faster and skips the web-page step, which is why it is not used. The database does not roll back with the artifact; its rollback is manual, from the snapshot the release command takes before it migrates a shared environment, and a restore creates a new database rather than overwriting the old.

The response-time and cadence figures are properties of the application, not of the deployment model. The application is one of the plain stacks measured in the first author's tokens-to-trace study~\cite{TokensToTrace}, server code that talks to the database directly and web pages without a framework, and the authors' view is that this is where those figures come from; that claim is not tested here.

The finding the authors would state first to a team adopting the model is the uncovered release units: draw the promotion boundary around everything a release changes, and treat a unit the pipeline does not carry as a half release until it does. The second is the control table: the promotion boundary is where the checks go, and drawing the boundary without the checks gives the audit trail and the rollback, not the assurance.

\section{Limitations and Threats to Validity}

Both authors built and operate the pipeline and extracted every figure from their own accounts; the per-rollout timings are in the Appendix so that the summary can be checked. No Model~A deployment of this system was measured, so the comparison in the Discussion is reasoning from Part~I, not a result. The system was under no named framework and no audit; the control table is the authors' own reading. One application, one cloud provider, one region; multi-region rollouts, blue-green and canary overlays, and stateful-workload migrations are out of scope. The deployment timings are from two-server fleets under one rolling policy with default thresholds; a larger fleet or a different policy changes the platform's share of the time, not the artifact's. The service-level figures predate the pipeline on the environment they were taken from. The connection budget was computed, not observed at its bound. The pipeline has no signature check, no scan, and no gate between upload and production; the approval record lives outside it. Database changes run from the release command outside the autonomous pipeline; carrying them inside it raises correctness questions not addressed here. The regulatory section names the regimes and does not attempt the control-by-control mapping a compliance package requires.

\section{Conclusion}

A pipeline in which the artifact's path from upload to running fleet is autonomous can be operated reliably under two conditions: each server learns its environment from what it is permitted to read rather than from a setting placed by hand, and each server fetches its settings at start from one per-environment store. Part~I argued that artifact promotion is the right control model. Part~II reports that the mechanism is small, that it carried 22 releases in its first month with a 36-second installation per server inside a six-minute rollout, that it draws the promotion boundary and places none of the checks on it, and that the two things it got wrong were both outside the boundary it drew: a release unit the pipeline did not promote, and a settings change that reaches a running fleet only by a redeploy.

\section*{Author Note}

The first author's prior work spans the regulated domains treated here. In a Canadian provincial health system he produced and distributed hashable binary packages, in C and Java, to hospitals, health-care facilities, and medical-education institutions under the province's health-information-protection law; the distributing organization required a fixed, hashable binary as a precondition. He also built financial rule engines, among them a provincial consolidation of financial and statistical accounts.

The second author implemented the deployment infrastructure of Part~II: the upload-triggered script, the managed platform environment, and the permission scoping. He holds a PhD and is a certified cloud- and information-security practitioner (CISSP, CCSP).

\appendix
\section{The 31 production rollouts}

Times in UTC; durations in seconds. Columns: script start; script start to platform update starting; first server installed; second server, from the first's completion; second server's completion to reported deployed; platform update, start to reported deployed; script, start to end.

\begin{center}
\footnotesize
\begin{tabular}{@{}rlrrrrrr@{}}
\toprule
\# & start & prep & server 1 & server 2 & settle & platform & script \\
\midrule
1 & 28 Jul 02:55 & 26 & 36 & 172 & 131 & 344 & 398 \\
2 & 28 Jul 03:11 & 30 & 36 & 172 & 141 & 354 & 402 \\
3 & 28 Jul 14:54 & 23 & 36 & 172 & 142 & 355 & 396 \\
4 & 29 Jul 01:31 & 26 & 37 & 166 & 141 & 349 & 398 \\
5 & 29 Jul 18:47 & 25 & 37 & 164 & 138 & 344 & 398 \\
6 & 31 Jul 21:06 & 25 & 36 & 151 & 128 & 320 & 366 \\
7 & 02 Aug 19:12 & 24 & 36 & 171 & 138 & 350 & 396 \\
8 & 04 Aug 03:42 & 24 & 36 & 170 & 137 & 349 & 396 \\
9 & 04 Aug 19:52 & 24 & 36 & 161 & 139 & 340 & 396 \\
10 & 08 Aug 21:57 & 24 & 36 & 163 & 128 & 332 & 366 \\
11 & 08 Aug 23:54 & 23 & 36 & 170 & 128 & 338 & 364 \\
12 & 09 Aug 02:11 & 34 & 36 & 171 & 137 & 349 & 406 \\
13 & 11 Aug 01:03 & 26 & 36 & 170 & 138 & 348 & 398 \\
14 & 11 Aug 19:57 & 29 & 36 & 164 & 138 & 344 & 401 \\
15 & 11 Aug 20:57 & 26 & 36 & 170 & 149 & 360 & 398 \\
16 & 12 Aug 19:04 & 28 & 36 & 160 & 138 & 339 & 369 \\
17 & 14 Aug 20:46 & 29 & 36 & 170 & 138 & 349 & 401 \\
18 & 14 Aug 21:16 & 28 & 36 & 161 & 130 & 330 & 370 \\
19 & 19 Aug 15:21 & 35 & 36 & 170 & 129 & 340 & 406 \\
20 & 19 Aug 18:08 & 32 & 36 & 170 & 139 & 350 & 404 \\
21 & 19 Aug 21:19 & 25 & 37 & 164 & 128 & 334 & 366 \\
22 & 20 Aug 15:50 & 30 & 38 & 163 & 138 & 344 & 402 \\
23 & 24 Aug 18:43 & 29 & 36 & 170 & 128 & 339 & 370 \\
24 & 25 Aug 01:35 & 27 & 36 & 170 & 138 & 348 & 401 \\
25 & 26 Aug 18:24 & 30 & 36 & 150 & 128 & 319 & 373 \\
26 & 27 Aug 04:29 & 33 & 36 & 171 & 131 & 342 & 406 \\
27 & 27 Aug 18:52 & 27 & 36 & 160 & 127 & 328 & 368 \\
28 & 30 Aug 22:57 & 39 & 36 & 170 & 138 & 349 & 411 \\
29 & 02 Sep 02:31 & 30 & 36 & 160 & 128 & 328 & 371 \\
30 & 02 Sep 02:41 & 11 & 36 & 159 & 128 & 328 & 352 \\
31 & 03 Sep 19:08 & 34 & 36 & 171 & 138 & 351 & 407 \\
\bottomrule
\end{tabular}
\end{center}

\end{document}